\documentclass[letterpaper, 10 pt, conference]{ieeeconf}  
\IEEEoverridecommandlockouts

\title{\LARGE \bf Formal Uncertainty Propagation for Stochastic Dynamical Systems with Additive Noise}

\author{Steven Adams\(^*\), Eduardo Figueiredo\(^*\)\thanks{\(^*\) Equal contribution} and Luca Laurenti
\thanks{All authors are with the Delft Center for Systems and Control, Technical University of Delft, 2628 CD Delft, The Netherlands.
        {Corresponding author: \tt\small s.j.l.adams@tudelft.com}}%
}

\usepackage{afterpage}
\usepackage{url}
\usepackage{multirow} 
\usepackage{booktabs} 
\usepackage{cite}
\usepackage{subcaption}

\usepackage{amsthm}
\usepackage{amsfonts,bm,mathtools}
\usepackage[ruled,linesnumbered, noend]{algorithm2e}

\newtheorem{theorem}{Theorem}
\newtheorem{proposition}[theorem]{Proposition}%
\newtheorem{example}{Example}

\newtheorem{assumption}{Assumption}
\newtheorem{remark}{Remark}
\newtheorem{problem}{Problem} 

\def\Prob{\mathbb{P}}
\def\ProbQ{\mathbb{Q}}

\def\sProb{{\mathcal{P}}}
\def\sDiscProb{{\mathcal{D}}}

\def\DiscProb{\mathbb{D}}
\def\sigmaAlgebra{{\mathcal{B}}}

\def\signature{{\Delta}}

\def\nDist{\mathcal{N}}

\def\wasserstein{{\mathbb{W}}}

\def\indicator{{\mathbf{1}}}

\def\diag{\mathrm{diag}}

\newcommand{\norm}[1]{\left\lVert#1\right\rVert}

\def\realNum{{\mathbb{R}}}
\def\natNum{\mathbb{N}}

\def\ball{{\mathbb{B}}}
\def\ambigSet{{\mathbb{S}}}
\def\sSimplex{\Pi}

\def\sA{{\mathcal{A}}}

\def\sC{{\mathcal{C}}}

\def\sL{{\mathcal{L}}}
\def\sM{{\mathcal{M}}}
\def\sN{{\mathcal{N}}}

\def\sP{{\mathcal{P}}}

\def\sR{{\mathcal{R}}}

\def\sX{{\mathcal{X}}}
\def\sY{{\mathcal{Y}}}

\def\bsC{{\bm{\mathcal{C}}}}

\def\bsR{{\bm{\mathcal{R}}}}

\def\mA{{{A}}}

\def\vzero{{{\bar{0}}}}
\def\vone{{{\bar{1}}}}

\def\vmu{{{\mu}}}
\def\vpi{{{\pi}}}

\def\vomega{{{\omega}}}

\def\vc{{{c}}}

\def\vm{{{m}}}

\def\vv{{{v}}}

\def\vx{{{x}}}

\def\vz{{{z}}}

\newcommand{\evpi}[1]{{\vpi^{(#1)}}}

\newcommand{\evx}[1]{{\vx^{(#1)}}}

\usepackage{xcolor}

\newcommand{\EF}[1]{{\color{blue}[EF: #1]}}

\begin{document}

\maketitle
\thispagestyle{empty}
\pagestyle{empty}

\begin{abstract}
In this paper, we consider discrete-time non-linear stochastic dynamical systems with additive process noise in which both the initial state and noise distributions are uncertain. Our goal is to quantify how the uncertainty in these distributions is propagated by the system dynamics for possibly infinite time steps. In particular, we model the uncertainty over input and noise as ambiguity sets of probability distributions close in the \(\rho\)-Wasserstein distance and aim to quantify how these sets evolve over time. Our approach relies on results from quantization theory, optimal transport, and stochastic optimization to construct ambiguity sets of distributions centered at mixture of Gaussian distributions that are guaranteed to contain the true sets for both finite and infinite prediction time horizons. We empirically evaluate the effectiveness of our framework in various benchmarks from the control and machine learning literature, showing how our approach can efficiently and formally quantify the uncertainty in linear and non-linear stochastic dynamical systems.   
\end{abstract}


\section{Introduction}\label{sec:intro}
Modern control systems are commonly uncertain. This uncertainty is due not only to the inherently stochastic dynamics \cite{stengel1994optimal}, but also to the use of data-driven methods and statistical estimators \cite{ljung2010perspectives, williams1995gaussian}.
The models of such systems are consequently not only stochastic, but the probability distributions of the various random variables are themselves uncertain \cite{van2015distributionally}. 
In many applications, where failures may have catastrophic consequences, these uncertainties cannot be ignored but must be formally accounted for during the temporal evolution of the system. 
Unfortunately, propagating a set of distributions through non-linear stochastic dynamics is computationally intractable \cite{landgraf2023probabilistic}.
This leads to the main question in this paper: Can we efficiently estimate the state of a non-linear stochastic dynamical system with uncertain initial and noise distributions over a possibly infinite time horizon while providing formal guarantees of correctness?

Various recent works have considered formal uncertainty propagation for dynamical systems. 
The more restrictive, but already intractable, problem of propagating known distributions over deterministic non-linear dynamics \cite{landgraf2023probabilistic} has been well studied, using techniques such as moment matching \cite{deisenroth2011pilco}, discretization-based methods \cite{julier2004unscented, arulampalam2002tutorial} or approximate numerical integration \cite{dunik2020state,ito2000gaussian}. 
However, these techniques commonly lack formal guarantees or are computationally demanding due to the need to discretize the full state space, and, critically, do not account for the fact that the actual underlying (state or noise) distributions are rarely known exactly.
The problem becomes even more challenging for uncertain distributions, as it requires propagating a set of distributions through the system dynamics. 
While this setting is receiving increasing attention, most existing approaches are either restricted to linear dynamics, lack formal guarantees, or do not scale to large prediction horizons \cite{aolaritei2022distributional,figueiredo2024uncertainty,figueiredo2025efficient}. 
For example, \cite{aolaritei2022distributional} proposes a framework for propagating sets of distributions close in Wasserstein distance through deterministic dynamics over multiple time steps, but the method is effectively limited to linear systems due to numerical tractability issues.
In the stochastic setting, \cite{figueiredo2024uncertainty} approximate the evolving state distributions with mixture distributions, while providing guarantees of correctness in the Total Variation metric. However, the resulting bounds generally become uninformative for small time horizons. 
Recently, \cite{figueiredo2025efficient} proposed propagating an uncertain distribution through non-linear dynamics by approximating it with discrete distributions with quantification of the resulting uncertainty in Wasserstein distance. While the method supports general stochastic dynamics, it requires constructing Cartesian products of state- and noise discretizations, limiting scalability.

In this paper, we present an algorithmic framework for formally propagating an uncertain input distribution through a non-linear dynamical system with possibly uncertain additive noise over a possibly infinite time horizon. Our approach relies on selecting a reference distribution from the input set, which is then propagated through the stochastic dynamics by approximating it with a mixture of Gaussian distributions. The resulting distribution is then used to construct a tractable convex set that is guaranteed to contain all possible distributions after propagation. 
To quantify uncertainty, we rely on the Wasserstein distance \cite{villani2008optimal}, which not only guarantees that distributions close in this distance have similar moments, but also enables the use of results from optimal transport \cite{villani2008optimal, peyre2019computational}. 
These results allow us to build a framework that not only efficiently propagates formal uncertainty sets but, for contracting systems, is even suitable for infinite-horizon propagation, as the radii of the resulting sets converge to a fixed point. 
We empirically illustrate the efficacy of our framework in various benchmarks. The experiments illustrate how our approach is able to successfully propagate uncertainty in various non-linear dynamical systems.

In summary, the main contributions of this work are:
\begin{itemize}
\item an algorithmic framework for obtaining tractable relaxations of sets of distributions propagated through non-linear dynamics with uncertain additive noise, with guarantees in Wasserstein distance, 
\item a proof that the Wasserstein bounds converge to a fixed point for contracting dynamics,
\item extensive empirical validation of our approach on several benchmarks, including neural network dynamics with multiple hidden layers. 
\end{itemize}




\section{Notation}
For a vector \(x \in \realNum^n\) we use \(\evx{i}\) to denote the \(i\)-th element of \(\vx\). Furthermore, for region \(\sX\subset\realNum^n\), the indicator function for $\sX$ is denoted as
\(
\indicator_{\sX}(x) = 
\begin{cases}
    1 & \text{if } x \in \sX\\
    0,              & \text{otherwise}
\end{cases}.
\) 
Given a measurable space \((\sX, \sigmaAlgebra(\sX))\) with \(\sigmaAlgebra(\sX)\) being the \(\sigma\)-algebra, we denote by \(\sProb(\sX)\) the set of probability distributions on \((\sX,\sigmaAlgebra(\sX))\).  The set of probability distributions with finite moments up to order \(\rho\in\natNum_{>0}\), \(\sProb_\rho(\sX)\), is defined as the set of all \(\Prob\in\sProb(\sX)\) such that \(\int_{\sX}\|\vx\|^\rho\Prob(d\vx)<\infty\), where \(\|\cdot\|\) is the \(L_\rho\)-norm. In this paper, for a metric space \(\sX\), \(\sigmaAlgebra(\sX)\) is assumed to be the Borel \(\sigma\)-algebra of \(\sX\). 
For a random variable \(\vx\) taking values in \(\sX\), \(\Prob_{\vx} \in \sProb(\sX)\) represents the probability measure associated to \(\vx\). For \(\rho\in\natNum_{>0}\) and \(\Prob,\ProbQ\in\sProb_\rho(\sX)\), the \emph{\(\rho\)-Wasserstein distance} between \(\Prob\) and \(\ProbQ\) is defined as:
    \begin{equation}
        \wasserstein_\rho(\Prob,\ProbQ)= \left(\inf_{\gamma\in\Gamma(\Prob,\ProbQ)}\int_{\sX\times\sX}\|\vx-\vx'\|^\rho\gamma(d\vx,d\vx')\right)^{\frac{1}{\rho}}
    \end{equation}
    where \(\Gamma(\Prob,\ProbQ)\subset\sProb(\sX\times\sX)\) represents the set of probability distributions with marginal distributions \(\Prob\) and \(\ProbQ\). For $\Prob \in \sP_\rho(\sX)$ and $\theta \geq 0$ the set of distributions closer than $\theta$ to $\Prob$ in the $\rho$-Wasserstein distance, also called \emph{\(\rho\)-Wasserstein ambiguity set}, is denoted by 
\begin{equation}
    \ball_\theta(\Prob) = \left\{\ProbQ\in\sProb_\rho(\sX)\mid\wasserstein_\rho(\Prob,\ProbQ)\leq\theta\right\}\subset\sProb_\rho(\sX).
\end{equation}

For \(N\in\natNum\), \(\sSimplex^N = \{ \pi \in \realNum^{N}_{\geq 0} \; : \; \sum_{i=1}^{N} \evpi{i} = 1 \}\) is the \(N\)-simplex. A discrete probability distribution \(\DiscProb\in \sProb(\sX)\) is defined as \(\DiscProb=\sum_{i=1}^N\evpi{i}\delta_{\vc_i}\), where \(\delta_\vc\) is the Dirac delta function centered at location \(\vc \in \sX\) and \(\pi \in \sSimplex^N\). The set of discrete probability distributions on \(\sX\) with at most \(N\) locations is denoted as \(\sDiscProb_N(\sX)\subset \sProb(\sX)\).  

For measurable spaces \((\sX,\sigmaAlgebra(\sX))\) and \((\sY,\sigmaAlgebra(\sY))\), a probability distribution \({\Prob} \in \sProb(\sX)\) and a measurable function \(g: \sX \rightarrow \sY \subseteq \realNum^{q}\), we use \(g\#\Prob\) to denote the push-forward measure of \(\Prob\) by \(g\), i.e., the measure on \(\sY\) such that for all \(\sA\in\sigmaAlgebra(\sY)\), \((g \# \Prob)(\sA) = \Prob(g^{-1}(\sA))\).  
For probability distributions \(\Prob,\ProbQ\in\sProb(\sX)\), we use \(\Prob\ast\ProbQ\in\sProb(\sX)\) to denote the convolution of \(\Prob\) and \(\ProbQ\), i.e., \((\Prob\ast\ProbQ)(\sA)=\int_{\sX\times\sX}\indicator_\sA(\vx+\vx')\Prob(d\vx)\ProbQ(d\vx')\) for all \(\sA\in\sigmaAlgebra(\sX)\).


\section{Problem Formulation}\label{sec:prob-form}
We consider a non-linear stochastic process with additive noise described as:
\begin{equation}\label{eq:system_dynamics}
    \vx_{k+1} = f(\vx_k) + \vomega_k, \quad \vx_0\sim\Prob_{\vx_0}, \vomega_k\sim\Prob_{\vomega}
\end{equation}
where, for state space \(\sX \subseteq \realNum^n\), \(f:\sX\mapsto\sX\) is a possibly non-linear measurable piecewise Lipschitz continuous function representing the one-step dynamics of System~\ref{eq:system_dynamics}. 
The initial distribution \(\Prob_{\vx_0}\in \sProb(\sX)\) and the process noise distribution \(\Prob_{\vomega}\in\sProb(\sX)\) are assumed to be unknown, as formalized in Assumption \ref{Assumption:Wasserstein} below. Intuitively, System~\ref{eq:system_dynamics} represents a general model of a \(n\)-dimensional discrete-time autonomous stochastic system with additive noise, but possibly non-linear dynamics, in which the knowledge of the probability distributions is uncertain. As a consequence, System \ref{eq:system_dynamics} encompasses a large class of modern control systems commonly used in practice, such as Neural Network dynamical systems \cite{nagabandi2018neural, adams2022formal}, or systems with additive noise learned from data \cite{mohajerin2018data, van2015distributionally}. 
\begin{assumption}
\label{Assumption:Wasserstein}
We assume that \(\Prob_{\vx_0}\) and \(\Prob_{\vomega}\) are uncertain. In particular, for $\rho\in \mathbb{N}_{>0}$, \(\theta_{\vx_0}, \theta_{\omega} \geq 0\), and known mixture of Gaussian distributions \(\bar\Prob_{\vx_0}, \bar\Prob_{\omega} \in \sP_{\rho}(\sX)\), we assume that \(\Prob_{\vx_0}\in \ball_{\theta_{\vx_0}}(\bar\Prob_{\vx_0})\)  and \(\Prob_{\vomega}\in  \ball_{\theta_{\omega}}(\bar\Prob_{\omega})\). That is, \(\Prob_{\vx_0}\) and \(\Prob_{\vomega}\) are only known to lie in certain \(\rho\)-Wasserstein ambiguity sets.
\end{assumption}
Assumption \ref{Assumption:Wasserstein} guarantees that we can quantify the uncertainty in System \eqref{eq:system_dynamics} using the $\rho-$Wasserstein distance.  This ensures that the resulting ambiguity sets could be estimated from data using existing data-driven approaches that guarantee closeness in the $\rho-$Wasserstein distance \cite{fournier2015rate, mohajerin2018data}. Furthermore, one can rely on the various properties of the $\rho-$Wasserstein distance, such as the fact that in contrast to other metrics commonly used such as KL-divergence \cite{gibbs2002choosing}, closeness in the $\rho-$Wasserstein distance guarantees closeness in the first $\rho-$moments \cite{adams2024finite}. We should also stress that the assumption that $\ball_{\theta_{\vx_0}}(\bar\Prob_{\vx_0})$ and $\ball_{\theta_{\omega}}(\bar\Prob_{\omega})$ are centered in mixtures of Gaussian distributions is not limiting, as this class of distributions can approximate any distribution arbitrarily well\footnote{The space of mixtures of Gaussians is dense in \(\sP_\rho(\sX)\) for the metric \(\wasserstein_\rho\) \cite{villani2008optimal, delon2020wasserstein}.} and the radius around it allows for distributions that are not mixtures nor Gaussian to be in the ambiguity sets.

Further, we note that if $\Prob_{\vx_{0}}$ and $\Prob_{\omega}$ were known,  then the distribution of the system at time $k$, $\Prob_{\vx_{k}}$, could be computed recursively as:
\begin{align}
\label{eq:TrueDynPropagation}
\Prob_{\vx_{k}}=(f\#\Prob_{\vx_{k-1}}) \ast \Prob_{\vomega},
\end{align}
that is, $\Prob_{\vx_{k}}$ is obtained by first propagating the distribution at the previous time step through state dynamics \(f\) and then taking the convolution with the noise distribution, which arises because of the additivity of the noise \cite{bogachev2007measure}. However, in System \eqref{eq:system_dynamics} both $\Prob_{\vx_{0}}$ and $\Prob_{\omega}$ are uncertain. Furthermore, we should emphasize that, in practice, even for known distributions,  the computation of Eqn \eqref{eq:TrueDynPropagation} in closed form is generally infeasible \cite{landgraf2023probabilistic}. This makes the setting considered in this paper particularly challenging.

\subsection{Problem Statement}
\begin{figure}[h]
    \centering
    \includegraphics[width=0.4\textwidth]{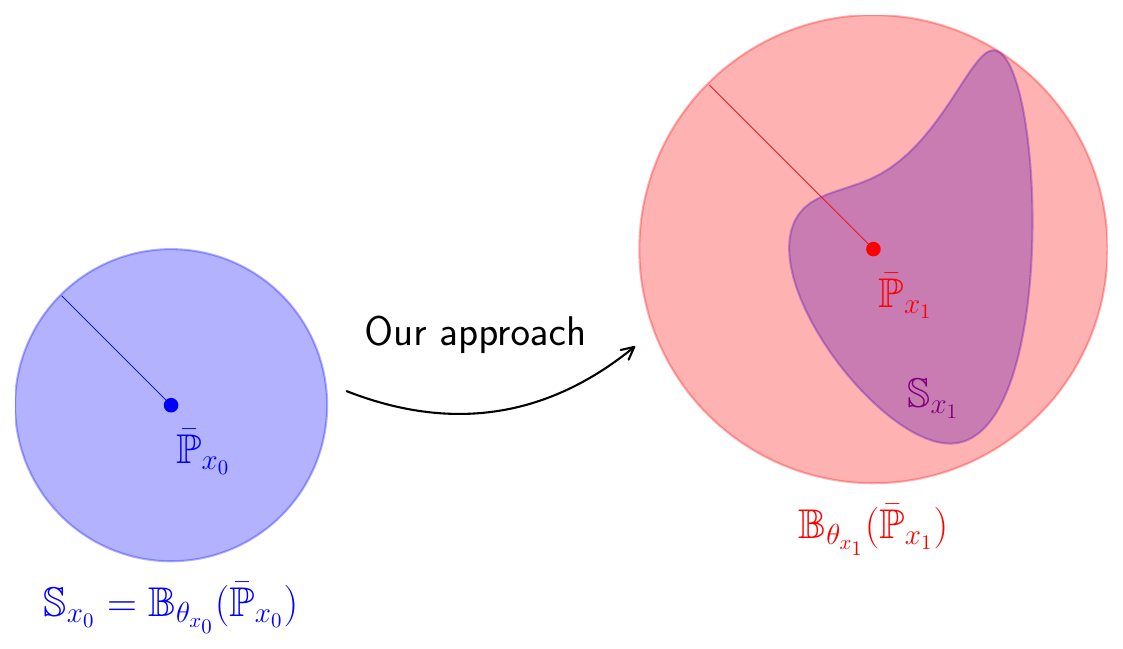}  
    \caption{Schematic representation of our proposed approach for $k=1$. As $\ambigSet_{\vx_1}$ is generally intractable, we over-approximate it with a \(\rho\)-Wasserstein ambiguity ball centered at
    $\bar{\Prob}_{\vx_{1}} \approx (f\#\bar{\Prob}_{\vx_{0}}) \ast \bar\Prob_{\vomega}$, that is, at the propagation of the center of the ambiguity ball at the previous time step through the system dynamics according to Eqn \eqref{eq:TrueDynPropagation}. The error in this approximation is formally accounted for in the choice of $\theta_{\vx_1}$.}
    \label{fig:ball-example}
\end{figure}
Given the ambiguity sets of initial and noise distributions introduced in Assumption \ref{Assumption:Wasserstein}, our goal is to quantify how this uncertainty is propagated by System \eqref{eq:system_dynamics} over time. In particular, we consider the following problem.
\begin{problem}\label{prob:main}
    Given time horizon \(K\in\natNum \cup \{\infty\}\), compute sets $\ambigSet_{\vx_0},...,\ambigSet_{\vx_K}$ defined recursively as follows for $k < K$:
    \begin{align}
        &  \ambigSet_{\vx_0}=\ball_{\theta_{\vx_0}}(\bar\Prob_{\vx_0}),  \label{eq:prob_ball_xt} \\
        &   \ambigSet_{\vx_{k+1}}=\{(f\#\Prob_{\vx_{k}}) \ast \Prob_{\vomega_k} \mid \Prob_{\vx_{k}}\in\ambigSet_{\vx_{k}}, \Prob_{\omega_k} \in \ball_{\theta_\vomega}(\bar\Prob_\vomega)\}.\nonumber
    \end{align}
\end{problem}
Following Eqn \eqref{eq:TrueDynPropagation},
sets $\ambigSet_{\vx_k}$ represent all the state distributions that System \ref{eq:system_dynamics} can assume at time step \(k\). Unfortunately, due to the potential nonlinearity of \(f\), solving Problem~\ref{prob:main} exactly is generally infeasible\footnote{Note that given a convex set $S$, and a non-linear map $T:S \mapsto V$, $T(S) \subset V$ is not necessarily convex \cite{boyd2004convex}. Consequently, sets $ \ambigSet_{\vx_k}$ are generally non-convex.}. Consequently, in what follows, we aim to design \(\rho\)-Wasserstein ambiguity sets of distributions that are guaranteed to include the true set of distributions of System \ref{eq:system_dynamics} over time.
We should emphasize that a solution of Problem \ref{prob:main} would allow one not only to formally quantify the uncertainty in dynamical systems and to compute worst-case expectations of various quantities due to the properties of the $\rho$-Wasserstein distance (see, e.g., Theorem 1 in \cite{gao2023distributionally}), but it would also represent an important step towards developing distributionally robust control algorithms for non-linear stochastic systems.

\begin{remark}
    If both the initial distribution \(\Prob_{\vx_0}\) and process noise \(\Prob_{\vomega}\) are known (i.e., \(\ambigSet_{\vx_0}=\{\Prob_{\vx_0}\}\) and  \(\ambigSet_{\vomega}=\{\Prob_{\vomega}\}\)), Problem~\ref{prob:main} reduces to compute Eqn \eqref{eq:TrueDynPropagation}, which, as mentioned earlier, is also object of research (e.g. \cite{figueiredo2024uncertainty}).
\end{remark}

\paragraph{Approach}
For every $k\leq K$, our approach is illustrated in Figure \ref{fig:ball-example} and is based on constructing center distributions \(\bar\Prob_{\vx_k} \subset \sP_\rho(\sX)\) and radius \(\theta_{\vx_k} \geq 0\) such that \(\ambigSet_{\vx_k} \subseteq \ball_{\theta_{\vx_k}}(\bar\Prob_{\vx_k})\) for any \(k \in \{1,\cdots,K\}\). More specifically, \(\bar\Prob_{\vx_k}\) is obtained in Section \ref{subsec:def-center} by approximating the push-forward distribution of the previous center by the dynamics of System \eqref{eq:system_dynamics} with a mixture of distributions. Then, in Section \ref{subsec:def-radii}, we show how 
a radius guaranteeing enclosure can be computed by relying on recent results on uncertainty propagation in non-linear stochastic systems \cite{figueiredo2025efficient}. Convergence guarantees of our approach are provided in Section \ref{subsec:def-radii}. Lastly, in Section~\ref{sec:experiments}, we demonstrate the efficacy of our approach through numerical experiments.



\section{Preliminaries}\label{sec:preliminaries}
Our approach relies on quantization of distributions and mixture distributions, which we recall in this Section.
\subsection{Quantization of Probability Distributions}\label{subsec:quantization}
For \(\sX\subseteq\realNum^n\), a set of $N$-points \(\bsC=\{\vc_i\}_{i=1}^N\) in \(\sX\) called locations, 
let \(\bsR=\{\sR_i\}_{i=1}^N\) be the Voronoi partition of \(\sX\) w.r.t. \(\bsC\), i.e., for each \(i\), i.e., $
    \sR_i = \left\{\vx\in\sX\mid \|\vz-\vc_i\|\leq\|\vz-\vc_j\|,\forall i\neq j \right\}.$ Then, the quantization operator \(\signature_{\bsC}:\realNum^n\rightarrow\realNum^n\) is defined as \(\signature_{\bsC}(\vx) \coloneqq \sum_{i=1}^N\vc_i\indicator_{\sR_i}(\vx)\). Intuitively, the quantization operator maps any point within the region $\sR_i$ to the location $c_i$. Consequently, for any probability distribution $\Prob \in \sP(\sX)$, it holds that
$    \signature_{\bsC} \# \Prob = \sum_{i=1}^N \Prob(\sR_i) \delta_{c_i} \in \DiscProb_N(\sX).$ 
Here, $\signature_{\bsC} \# \Prob$ is called the \emph{discretization} or \emph{quantization} of $\Prob$, effectively converting a possibly continuous distribution into a discrete one with support of size $N$ based on the specified locations.

\begin{remark}\label{remark:closed_form_quantization}
    $ \signature_{\bsC} \# \Prob$ can be  computed particularly efficiently when the probability mass of each Voronoi region can be expressed in analytic form, typically via the distribution’s cumulative distribution function. For instance, univariate Gaussian allow exact computation over intervals, while multivariate Gaussians admit closed-form expressions over regions aligned with the eigenbasis of the covariance matrix \cite{adams2024finite} (consequently, the same also holds for mixtures of Gaussians for which all components satisfy the same property).
\end{remark}

\subsection{Mixture Distributions}
A mixture distribution of size \(M\in\natNum\), is a set of \(M\) distributions of the same type, also called components, averaged w.r.t. a probability vector \(\vpi\in\Pi_M\), also called the weights \cite{adler2009random}. 
The convolution of a discrete distribution \(\sum_{i=1}^N\evpi{i}\delta_{\vc_i}\) with distribution \(\Prob\) results in a mixture distribution with components \(\delta_{\vc_i}\ast\Prob\) and weights \(\vpi\), i.e., 
\(
    \big(\sum_{i=1}^N\evpi{i}\delta_{\vc_i}\big)\ast\Prob=\sum_{i=1}^N\evpi{i}(\delta_{\vc_i}\ast\Prob), 
\)
where \((\delta_{\vc}\ast\Prob)(d\vx)= \Prob(d\vx - \vc)\). 
When \( \Prob = \sN(m, \Sigma) \), i.e. a Gaussian distribution with mean \( m \) and covariance \(\Sigma\), \( \delta_{\vc_i}\ast\Prob = \sN(m+c_i, \Sigma) \) and, thus, \( \sum_{i=1}^N\evpi{i}(\delta_{\vc_i}\ast\Prob) = \sum_{i=1}^N\evpi{i} \sN(m+c_i, \Sigma) \) is a Gaussian Mixture.



\section{Approximation Scheme}\label{sec:approx-scheme}
In this section, we introduce our approach to solve Problem \ref{prob:main} by iteratively approximating the true sets of uncertain distributions \(\ambigSet_{\vx_1}, \hdots,\ambigSet_{\vx_K}\) with a set of \(\rho\)-Wasserstein ambiguity balls \(\ball_{\theta_1}(\bar\Prob_{\vx_1}), \hdots,\ball_{\theta_K}(\bar\Prob_{\vx_K})\).
We first show that due to the additive nature of the process noise in System~\ref{eq:system_dynamics} a natural choice is to consider centers \(\bar\Prob_{\vx_1}, \hdots,\bar\Prob_{\vx_K}\) as a mixture of Gaussian distributions 
approximating the pushforward distribution of the center at the previous time, that is, of the time evolution of $\bar{\Prob}_{x_0}$ through the dynamics in System \eqref{eq:system_dynamics}. After that, we determine the radii \(\theta_1,\hdots,\theta_K\) to ensure that \(\ball_{\theta_k}(\bar\Prob_{\vx_k})\) is guaranteed to include \(\ambigSet_{\vx_k}\) for at all time steps \(k\). Further, in Proposition \ref{prop:convergence-contractive}, we show that, for contractive systems, the radius of the ambiguity ball converges to a fixed finite value as $k\to\infty$. Lastly, in Subsection \ref{subsec:algorithm}, we present the full algorithm for our approximation scheme.

\subsection{Ambiguity Ball Centers via Mixture Distributions}\label{subsec:def-center}
For each time step \(k\), 
consider a set of locations \(\bsC_k\subset\sX\) that define a quantization operator \(\signature_{\bsC_k}\), where we refer to Subsection~\ref{subsec:algorithm} for how to select \(\bsC_k\). 
Given \(\bar\Prob_{\vx_0}\) and \(\bar\Prob_{\vomega}\), which, for clarity of presentation and without loss of generality,
we assume are Gaussian mixtures with equal covariance across components, we define the center of the ambiguity balls \(\bar\Prob_{\vx_{k+1}}\) iteratively for each \(k\in\{0,\hdots K-1\}\) as follows\footnote{In the general case of mixture of Gaussians with different covariance per component, following standard practice \cite{adams2024finite}, one can apply a separate quantization operator to each mixture component. That is, for \(\Prob=\sum_{i=1}^N\evpi{i}\Prob_i\), one can quantization locations \(\bsC_i\) for each \(\Prob_i\) and take \(\sum_{i=1}^N\evpi{i}\signature_{\sC_i}\#\Prob_i\) as the quantization of \(\Prob\).}:
\begin{align}
    \bar\Prob_{\vx_{k+1}} &= (f\#\signature_{\bsC_k}\#\bar\Prob_{\vx_{k}})\ast\bar\Prob_{\vomega} 
    \nonumber \\
    &=\sum_{i=1}^{|\bsC_k|}\bar\Prob_{\vx_{k}}(\sR_{k,i})(\delta_{f(\vc_{k,i})}\ast\bar\Prob_{\vomega}), 
    \label{eq:approx_centers}
\end{align}
where \(\sR_{k,i}\) is the region in the Voronoi partition of \(\sX\) w.r.t. \(\bsC_k\) that corresponds to location \(\vc_{k,i}\in \bsC_k\).
Note that, since all components of mixture \(\bar\Prob_{\vomega}\) share the same covariance, the components of the resulting mixtures \(\bar\Prob_{\vx_{k+1}}\) have identical covariances. As explained in Remark~\ref{remark:closed_form_quantization}, this choice guarantees that \(\bar\Prob_{\vx_{k}}(\sR_{k,i})\) can be computed in closed form for a suitable choice of locations \(\bsC_k\), thereby facilitating efficient quantization in Subsection~\ref{subsec:algorithm}. 
Intuitively, at each time \(k\), Eqn~\eqref{eq:approx_centers} builds a mixture approximation of \((f\#\bar\Prob_{\vx_{k}})\ast\bar\Prob_{\vomega}\) by averaging and shifting \(\bar\Prob_\vomega\) with the weights and locations resulting from the quantization of \(\bar\Prob_{\vx_{k}}\) by \(\signature_{\bsC_k}\). 
Note that the mixture approximation consists of a straightforward application of state dynamics \(f\) to the support of the discrete approximation of the state dynamics at the previous time step, i..e,  \(\vc_{k,1},\hdots\vc_{k,|\bsC_k|}\). 
\begin{example}
    Consider \(\bar\Prob_\vomega\) to be a known Gaussian noise distribution with mean \(\vm\) and covariance \(\Sigma\), i.e., \(\bar\Prob_\vomega=\nDist(\vm,\Sigma)\). The center distributions defined by 
    Eqn~\eqref{eq:approx_centers} for each time step \(k\) are GMMs that can be written as
    \[
        \bar\Prob_{\vx_{k+1}}=\sum_{i=1}^{|\bsC_k|}\bar\Prob_{\vx_k}(\sR_{k,i})\nDist(f(\vc_{k,i})+\vm,\Sigma).
    \]
\end{example}

\subsection{The Radii of the Ambiguity Balls}\label{subsec:def-radii}
After defining the centers of the ambiguity balls, we now determine the radii \(\theta_k\). The following Proposition presents an upper bound for the radii that ensure that \(\ambigSet_{\vx_k}\subseteq\ball_{\theta_k}(\bar\Prob_{\vx_k})\) for all time steps \(k\leq K\).
\begin{proposition}\label{prop:theor_bound_theta}
    Let \(\bar\Prob_{\vx_1}, \hdots,\bar\Prob_{\vx_K}\) be defined according to Eqn~\eqref{eq:approx_centers}. For each \(k\in\{0,\hdots,K-1\}\), iteratively define 
    \begin{align}\label{eq:theta-condition-for-superset}
        \theta_{\vx_{k+1}} \geq \sup_{\substack{
            \Prob\in \ball_{\theta_{x_k}}(\bar\Prob_{x_k}), \\ \Prob_\omega \in \ball_{\theta_\omega}(\bar\Prob_\omega)}}\wasserstein_\rho(f\#\Prob\ast\Prob_\omega, \bar\Prob_{\vx_{k+1}})
    \end{align}    
    Then, $\forall k\in\natNum$ it holds that $
        \ambigSet_{\vx_{k}} \subseteq \ball_{\theta_{\vx_k}}(\bar\Prob_{\vx_{k}}).$
\end{proposition}
The proof of Proposition~\ref{prop:theor_bound_theta} is in Section~\ref{sec:proofs} and relies on the fact that, after selecting a particular center distribution \(\bar\Prob_{\vx_{k}}\), it is enough to take a radius larger than the worst distance of this center to a set of distributions containing the true ambiguity set $\ambigSet_{\vx_k}$.
While computing the supremum in Eqn \eqref{eq:theta-condition-for-superset} is generally intractable, upper bounds can be computed by extending results in \cite{figueiredo2025efficient} as we show in Proposition \ref{prop:main-bound-result}.
\begin{proposition}\label{prop:main-bound-result}
    Consider a set of locations $\bsC\subset\sX$. Further, call
     \begin{equation}\label{def:theta_d-definition}
         \theta_{\Delta} = \bigg( \sum_{\ell=1}^{|\bsC|} \int_{\sR_{\ell}} \norm{x-c_{\ell}}^\rho d\bar\Prob_{\vx_k}(\vx) \bigg)^\frac{1}{\rho},
     \end{equation}
     and for \(\ell \in\{1,\hdots,|\bsC|\}\), let $\alpha_{\ell}, \beta_{\ell} \in \realNum_{\geq0}$ be such that for $\vx\in \sX$
    \begin{equation}\label{eq:norm-linearization}
        \norm{f(\vx) - f(\vc_{\ell})}^\rho \leq \alpha_{\ell} \norm{x - \vc_{\ell}}^\rho + \beta_{\ell}.
    \end{equation}
    Then, for $\hat{\alpha} =\max_{\ell\in\{1,\hdots,|\bsC|\}}\alpha_{\ell}$, it holds that
    \begin{align}
        &\sup_{\substack{
            \Prob\in \ball_{\theta_{x_k}}(\bar\Prob_{x_k}), \\ \Prob_\omega \in \ball_{\theta_\omega}(\bar\Prob_\omega)}}\wasserstein_\rho(f\#\Prob\ast\Prob_\omega, \bar\Prob_{\vx_{k+1}}) \nonumber \\
        &\qquad \leq \theta_\omega + \bigg(\hat{\alpha}(\theta_{\vx_k}+\theta_{\Delta})^\rho + \sum_{\ell=1}^{|\bsC|}\bar\Prob_{\vx_k}(\sR_{\ell})\beta_{\ell}\bigg)^\frac{1}{\rho}.  \label{eq:bound-thm-3}
    \end{align}
\end{proposition}
Proposition \ref{prop:main-bound-result} guarantees that the radius \(\theta_{\vx_{k+1}}\) of our proposed ambiguity ball \(\ball_{\theta_{\vx_{k+1}}}(\bar\Prob_{\vx_{k+1}})\) can be taken as in Eqn \eqref{eq:bound-thm-3}. This bound crucially depends on \(\theta_\Delta\), which represents a quantization penalty as explained in Remark~3 from \cite{figueiredo2025efficient} and the norm linearization of \(f\) around each point \(\vc_i\) in Eqn \eqref{eq:norm-linearization}.
Note that both the center and the radius of the ambiguity balls depend on the choice of quantization locations \(\bsC_k\). The selection of the quantization operator, as well as the computation of the quantization penalty and the norm linearization, is discussed in the next subsection. 

Before defining an algorithm for Problem \ref{prob:main}, it is important to discuss how the radii of the ambiguity sets resulting from Proposition \ref{prop:main-bound-result} evolve over time. In Proposition \ref{prop:convergence-contractive} below, we show that when System \ref{eq:system_dynamics} is contractive (i.e. the global Lipschitz constant of $f$ is smaller than one, i.e. \(\sL_f < 1\)), the \(\rho\)-Wasserstein ambiguity radius in Eqn \eqref{eq:bound-thm-3} converges to a fixed point, which allows us to obtain informative (i.e. finite radius) sets also for arbitrarily large $K$.

\begin{proposition}\label{prop:convergence-contractive}
    Let $f:\sX \mapsto \sX$ be a piecewise Lipschitz continuous function with global Lipschitz constant $\sL_f<1$. Given $\epsilon > 0$, let $\bsC_k \in \sX$ be a set of points such that $\theta_{\Delta, k} = \bigg( \sum_{\ell=1}^{|\bsC_k|} \int_{\sR_{k, \ell}} \norm{x-c_{k, \ell}}^\rho d\bar\Prob_{\vx_k}(\vx) \bigg)^\frac{1}{\rho} \leq \epsilon$ for every $k$. Consider the following iterative process describing the evolution of the radius in Eqn \eqref{eq:bound-thm-3} for $k \in \natNum_{>0}$, with $\theta_0 = \theta_{\vx_0}$ and
    \begin{equation}
        \theta_{k+1} = \theta_\omega + \bigg(\hat{\alpha}_k(\theta_{k}+\theta_{\Delta, k})^\rho + \sum_{\ell=1}^{|\bsC_k|}\bar\Prob_{\vx_k}(\sR_{k, \ell})\beta_{k, \ell}\bigg)^\frac{1}{\rho}.
    \end{equation}
    Then,
    \begin{equation}
        \lim_{k\to\infty} \theta_k \leq \frac{\theta_\omega}{1-\sL_f} + \frac{\sL_f}{1-\sL_f}\epsilon.
    \end{equation}
\end{proposition}

Proposition \ref{prop:convergence-contractive} is similar to Theorem 7 in \cite{figueiredo2025efficient} and it relies on the Banach Fixed Point Theorem \cite{goebel1990topics}. The immediate consequence of this result is that for contractive systems, the radius of \(\ball_{\theta_{\vx_k}}(\bar\Prob_{\vx_k})\), \(\theta_{\vx_k}\), remains finite even for an infinite horizon. Also note that, while $\epsilon$ can be made arbitrarily small by controlling the quantization error, $\theta_w$ represents the uncertainty in the additive noise. Consequently, the radius of the $\rho-$Wasserstein ambiguity sets resulting from our approach will converge to $0$ only if the additive noise distribution is not uncertain. This is intuitive because while the uncertainty in the initial distribution can be made arbitrarily small over time by the contractive nature of the dynamics, the uncertainty in the noise is added to the system at every time step.

\subsection{Algorithm}
\label{subsec:algorithm}

We summarize our procedure to construct the \(\rho\)-Wasserstein ambiguity sets with System~\ref{eq:system_dynamics} in Algorithm \ref{alg:propagate-ball}. 
In line 2, given the center of the ambiguity set at time \(k\geq0\), \(\bar\Prob_{\vx_{k}}\), we select the quantization locations \(\bsC_k\) according to Algorithm~2 in \cite{adams2024finite}. The resulting Voronoi partition \(\{\sR_{k,i}\}_{i=1}^{|\bsC_k|}\) w.r.t. \(\bsC_k\), is guaranteed to align with the eigenbasis of the identical covariance shared by all components of \(\bar\Prob_{\vx_{k}}\). This structure enables tractable quantization of \(\bar\Prob_{\vx_{k}}\) and ensures a closed form expression for the quantization error \(\theta_\signature\) in Eqn.~\eqref{def:theta_d-definition}. 
In line 3, the quantization of \(\bar\Prob_{\vx_{k}}\) is propagated through the system dynamics, resulting in a discrete distribution \(\DiscProb_{k+1}\). To control the number of components of the mixture, and thus the computational complexity of the subsequent quantizations, we include an optional compression step in line 4. This step approximates \(\DiscProb_{k+1}\) by a discrete distribution with fewer elements. We implement this compression, referred to as \(\texttt{Compress}\), by applying \(N\)-means clustering to the support of \(\DiscProb_{k+1}\). The error introduced by the compression step is accounted for by computing the \(\rho\)-Wasserstein distance between two discrete distributions which can be solved using linear programming techniques, as shown in \cite{peyre2019computational}. 
In line 5, the center of the ambiguity set at time \(k+1\) is obtained by convoluting the compressed discrete distribution with \(\bar\Prob_\vomega\). Finally, the radius of the ambiguity set is computed according to Proposition~\ref{prop:main-bound-result}. The quantization error \(\theta_\signature\) is computed according to Corollary~10 in \cite{adams2024finite}, and the norm linearization in Eqn.~\eqref{eq:norm-linearization} is computed as in Section V.A of \cite{figueiredo2025efficient}, using bound propagation techniques \cite{mathiesen2022safety}. 

\RestyleAlgo{ruled}
\SetKwComment{Comment}{/* }{ */}
\begin{algorithm}[h]
\DontPrintSemicolon
\caption{Propagate $\rho$-Wasserstein ambiguity ball with System \ref{eq:system_dynamics}}\label{alg:propagate-ball}
\KwIn{Function $f$ from System \ref{eq:system_dynamics}, state and noise ambiguity balls, $\ball_{\theta_{\vx_k}}(\bar\Prob_{\vx_k}), \ball_{\theta_\omega}(\bar\Prob_\omega)$ at step $k$}
\KwOut{Ambiguity ball $\ball_{\theta_{\vx_{k+1}}}(\bar\Prob_{\vx_{k+1}})$ according to approximation scheme in Section \ref{sec:approx-scheme}}
\SetKwFunction{FMain}{PropagateAmbiguityBall}
\SetKwProg{Fn}{function}{:}{}
\Fn{\FMain{$f$, $\ball_{\theta_{\vx_k}}(\bar\Prob_{\vx_k})$, $\ball_{\theta_\omega}(\bar\Prob_\omega)$}}{
    $\bsC_{k} \gets \texttt{Algorithm 2 from \cite{adams2024finite}}$\;
    $\DiscProb_{k+1} \gets f\#\Delta_{\bsC_{k}}\#\bar\Prob_{x_k}$\;
    $\DiscProb_{k+1}, \theta_{\text{compr}}\gets \texttt{Compress}(\DiscProb_{\vx_{k+1}}, N)$\;
    $\bar\Prob_{\vx_{k+1}} \gets \DiscProb_{k+1}\ast\bar\Prob_\omega$\;
    $\theta_{\vx_{k+1}} \gets \texttt{Prop. \ref{prop:main-bound-result} with } \theta = \theta_{\vx_k}+\theta_{\text{compr}}$\;
    }
    \Return{$\ball_{\theta_{\vx_{k+1}}}(\bar\Prob_{\vx_{k+1}})$}
\end{algorithm}

\section{Experiments}\label{sec:experiments}
\begin{figure*}[t]
    \centering
    \vspace{-1em}
    \begin{subfigure}[t]{0.32\textwidth}
        \raisebox{0.1\height}{
        \includegraphics[width=\linewidth]{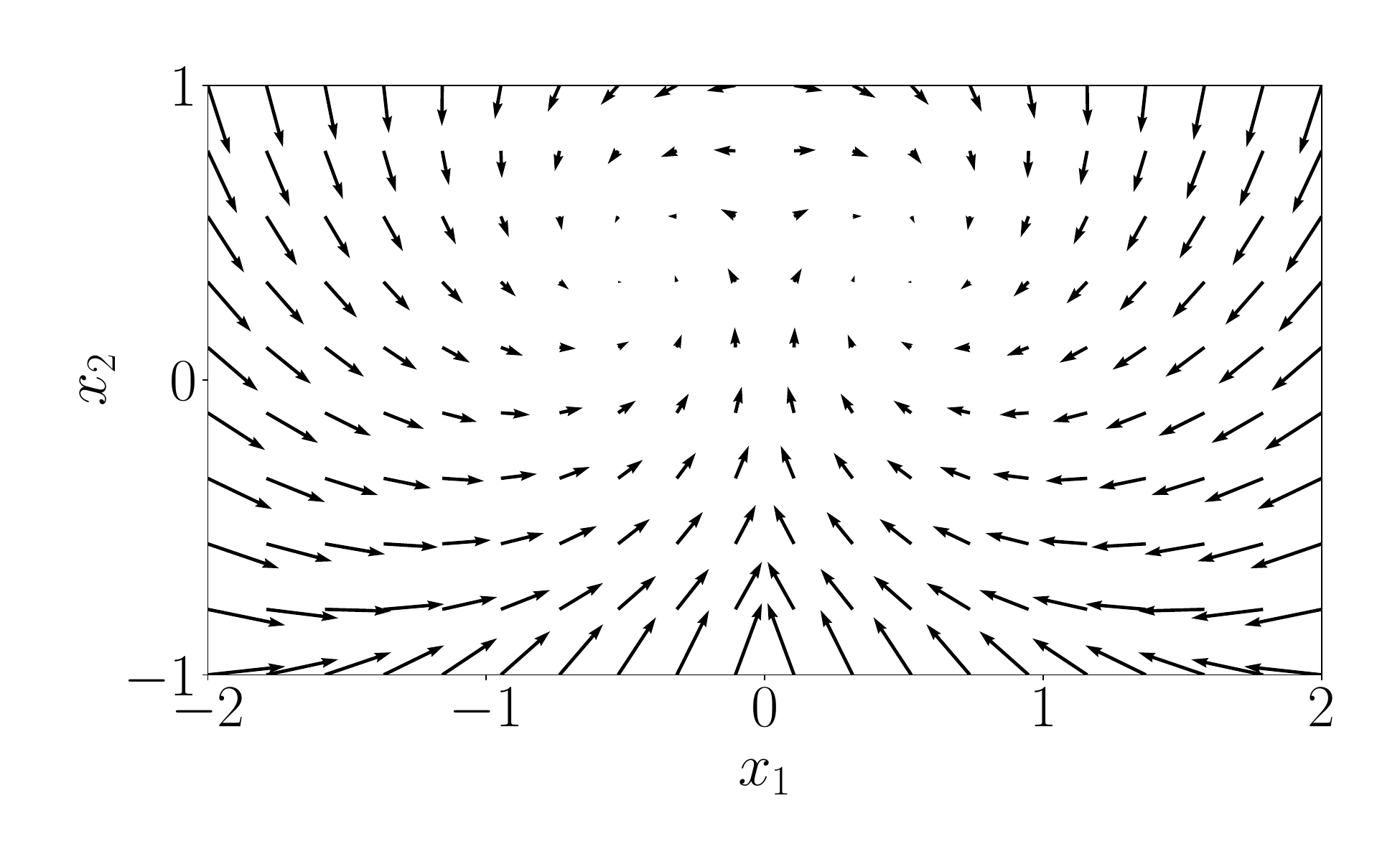}
        }
        \caption{Double Spiral, Vector Field}
        \label{fig:double_spiral_dynamics}
    \end{subfigure}
    \hfill
    \begin{subfigure}[t]{0.32\textwidth}
        \includegraphics[width=\linewidth]{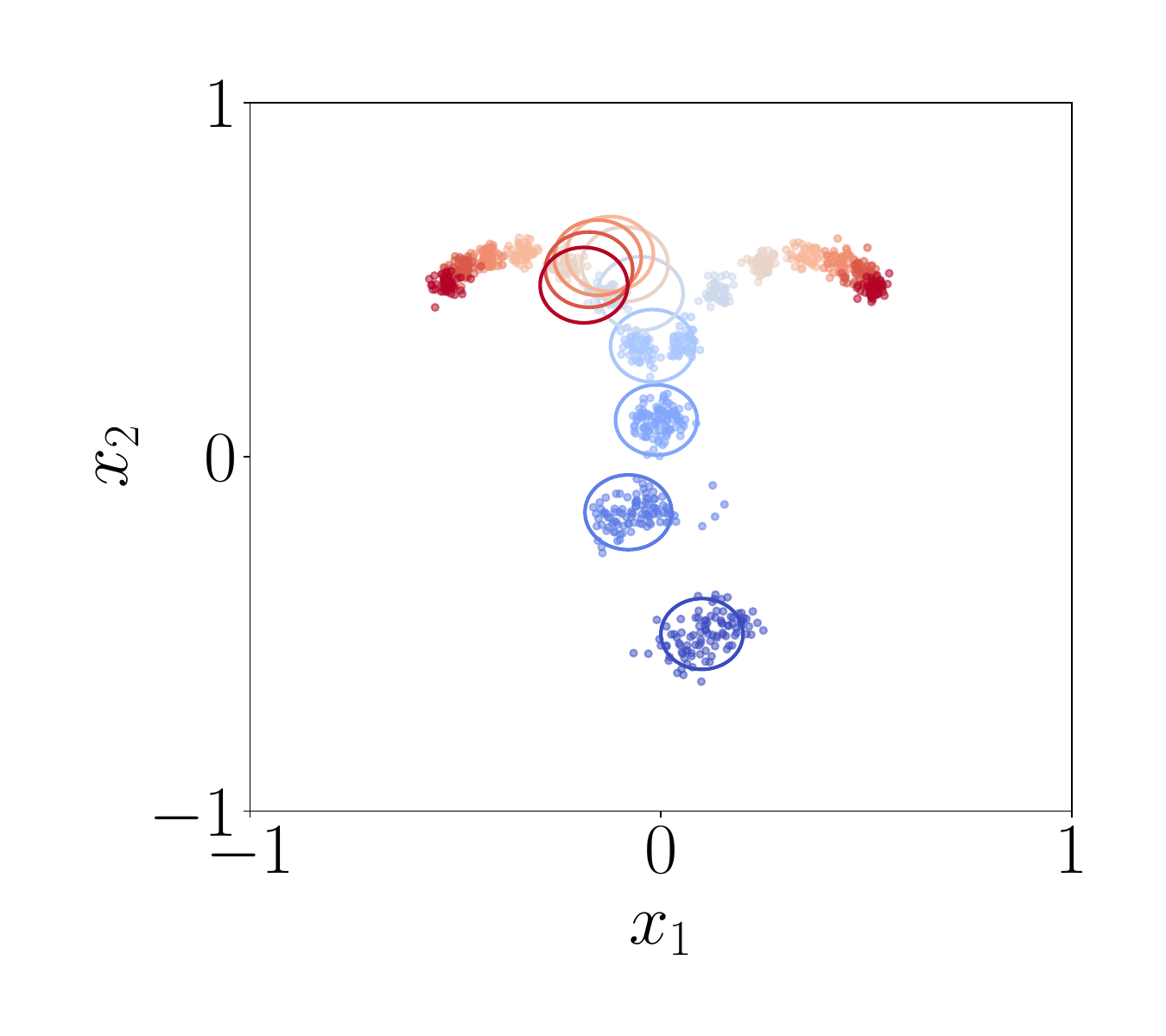}
        \caption{Double Spiral, True Distributions}
        \label{fig:double_spiral_true_dist}
    \end{subfigure}
    \hfill
    \begin{subfigure}[t]{0.32\textwidth}
        \includegraphics[width=\linewidth]{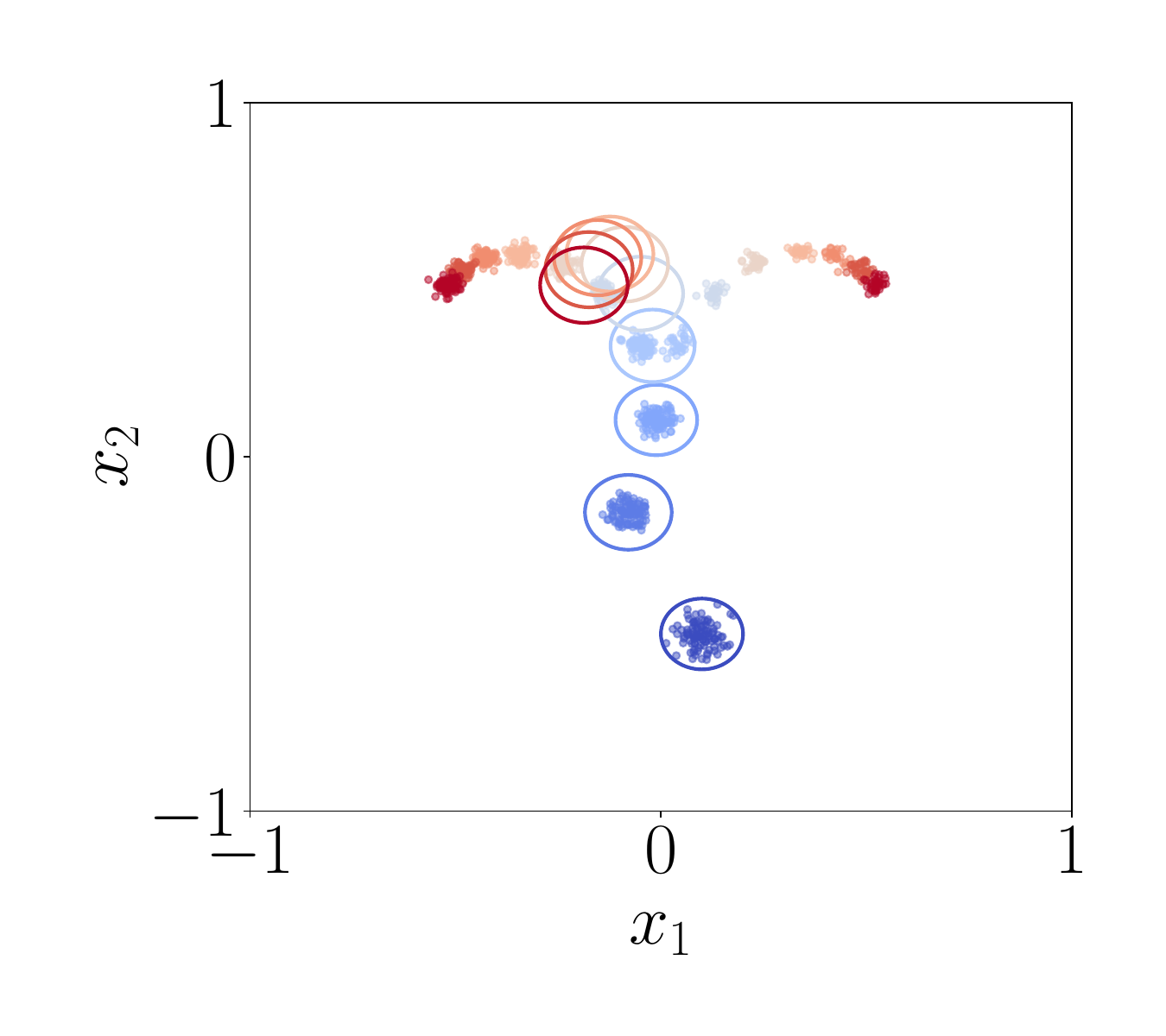}
        \caption{Double Spiral, Approx. Distribution}
        \label{fig:double_spiral_apporx_dist}
    \end{subfigure}
    \begin{subfigure}[t]{0.32\textwidth}
        \includegraphics[width=\linewidth]{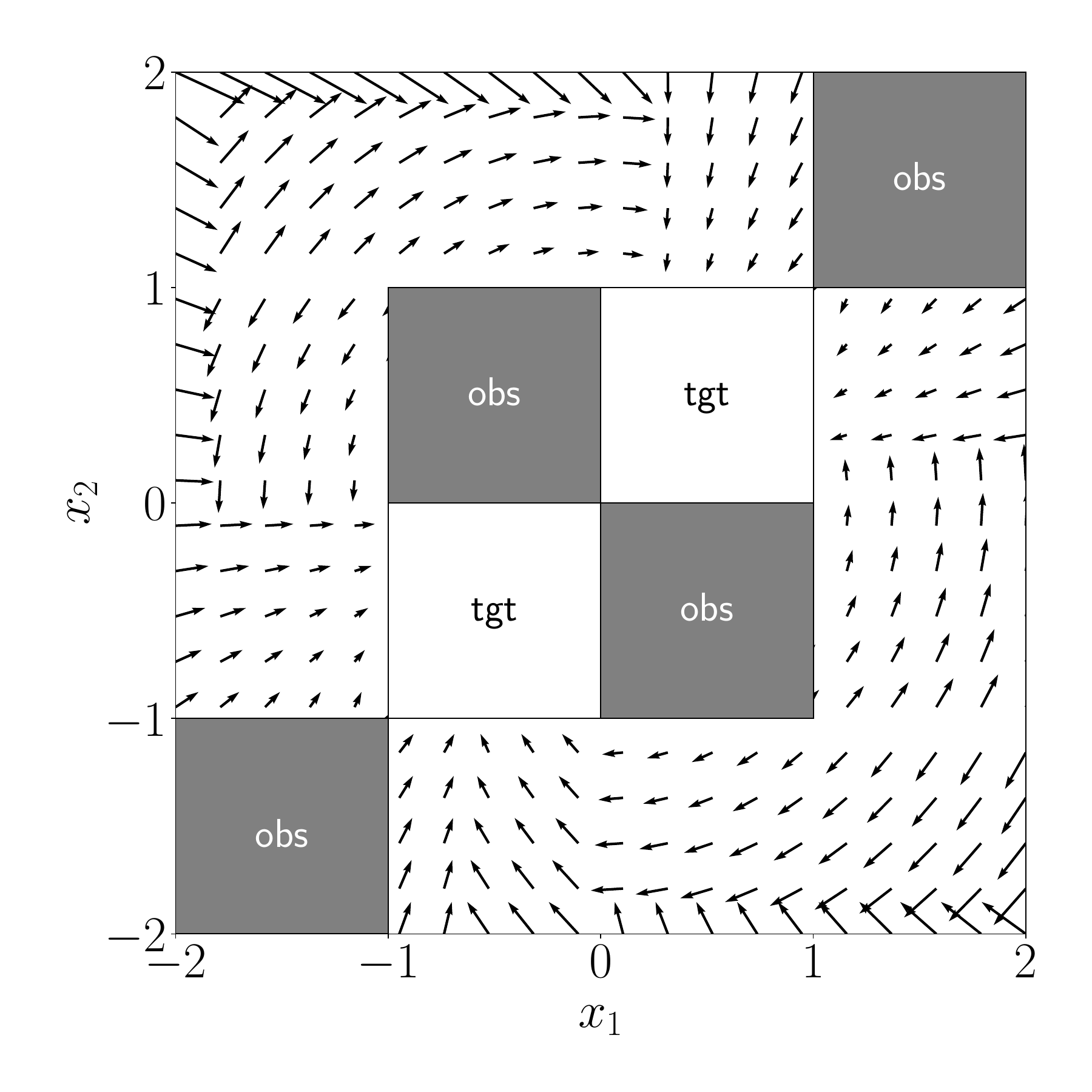}
        \caption{Piecewise Linear, Vector Field}
        \label{fig:switched_dynamics}
    \end{subfigure}
    \hfill
    \begin{subfigure}[t]{0.32\textwidth}
        \includegraphics[width=\linewidth]{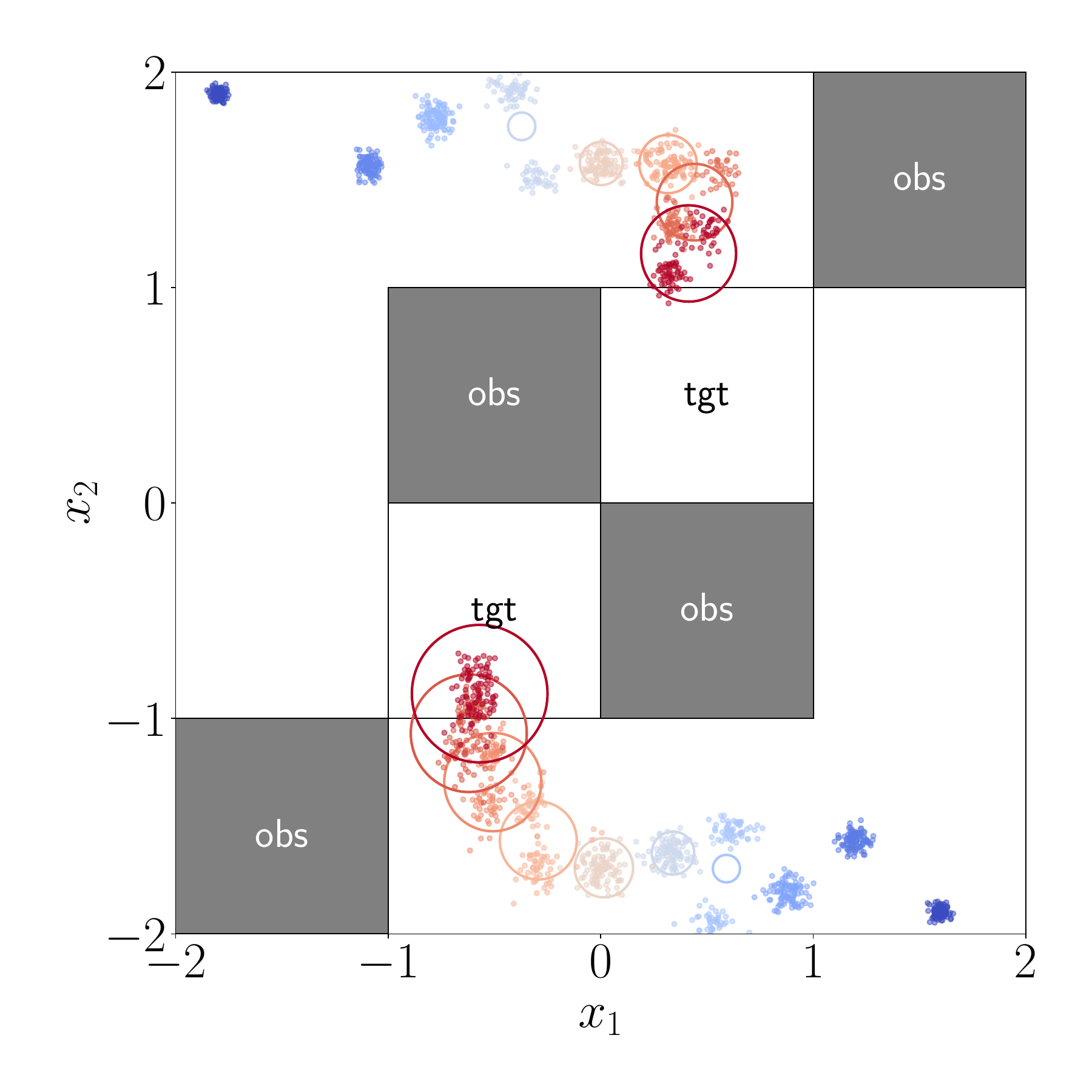}
        \caption{Piecewise Linear, True Distributions}
        \label{fig:switched_true_dist}
    \end{subfigure}
    \hfill
    \begin{subfigure}[t]{0.32\textwidth}
        \includegraphics[width=\linewidth]{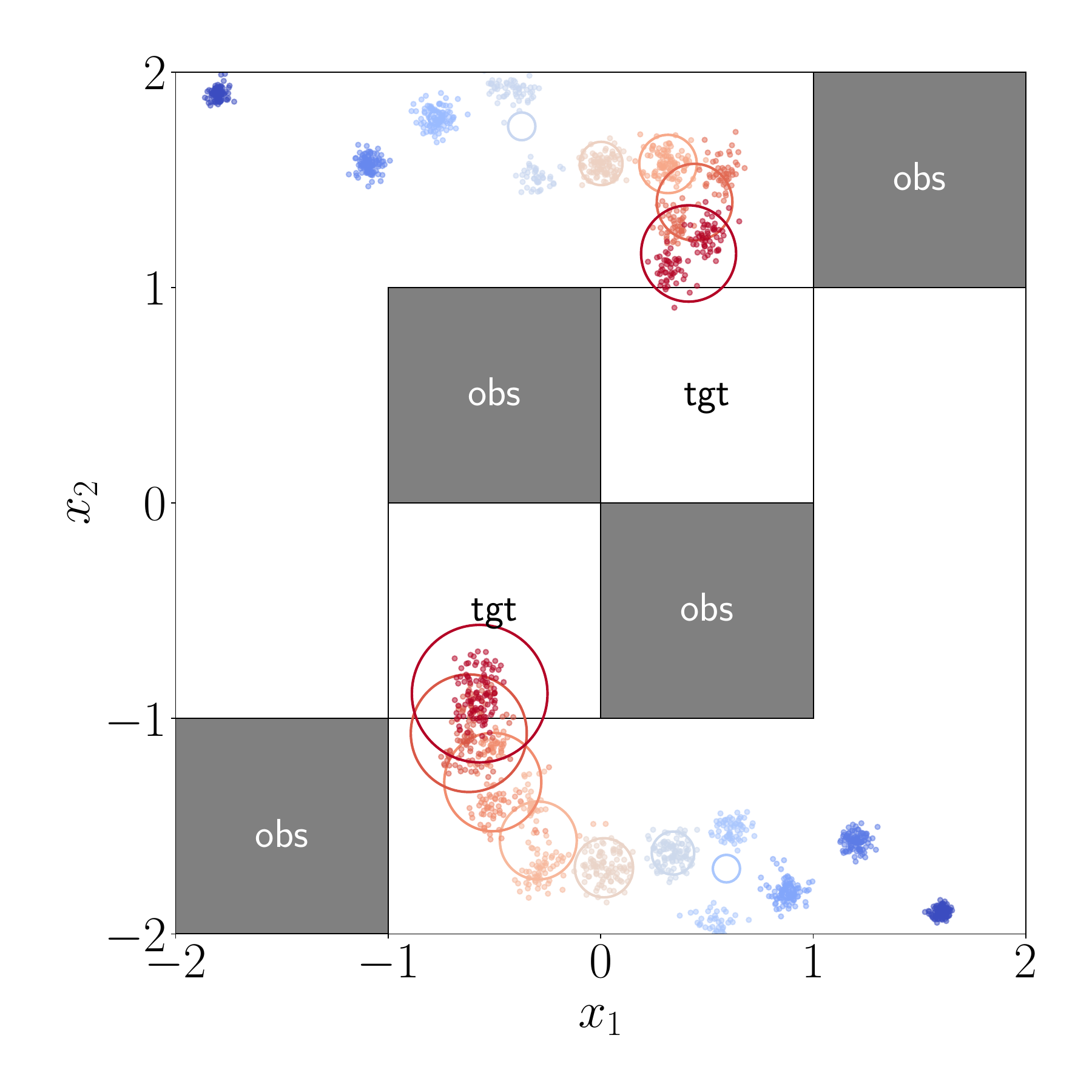}
        \caption{Piecewise Linear, Approx. Distribution}
        \label{fig:switched_approx_dist}
    \end{subfigure}
    \begin{subfigure}[t]{0.32\textwidth}
        \includegraphics[width=\linewidth]{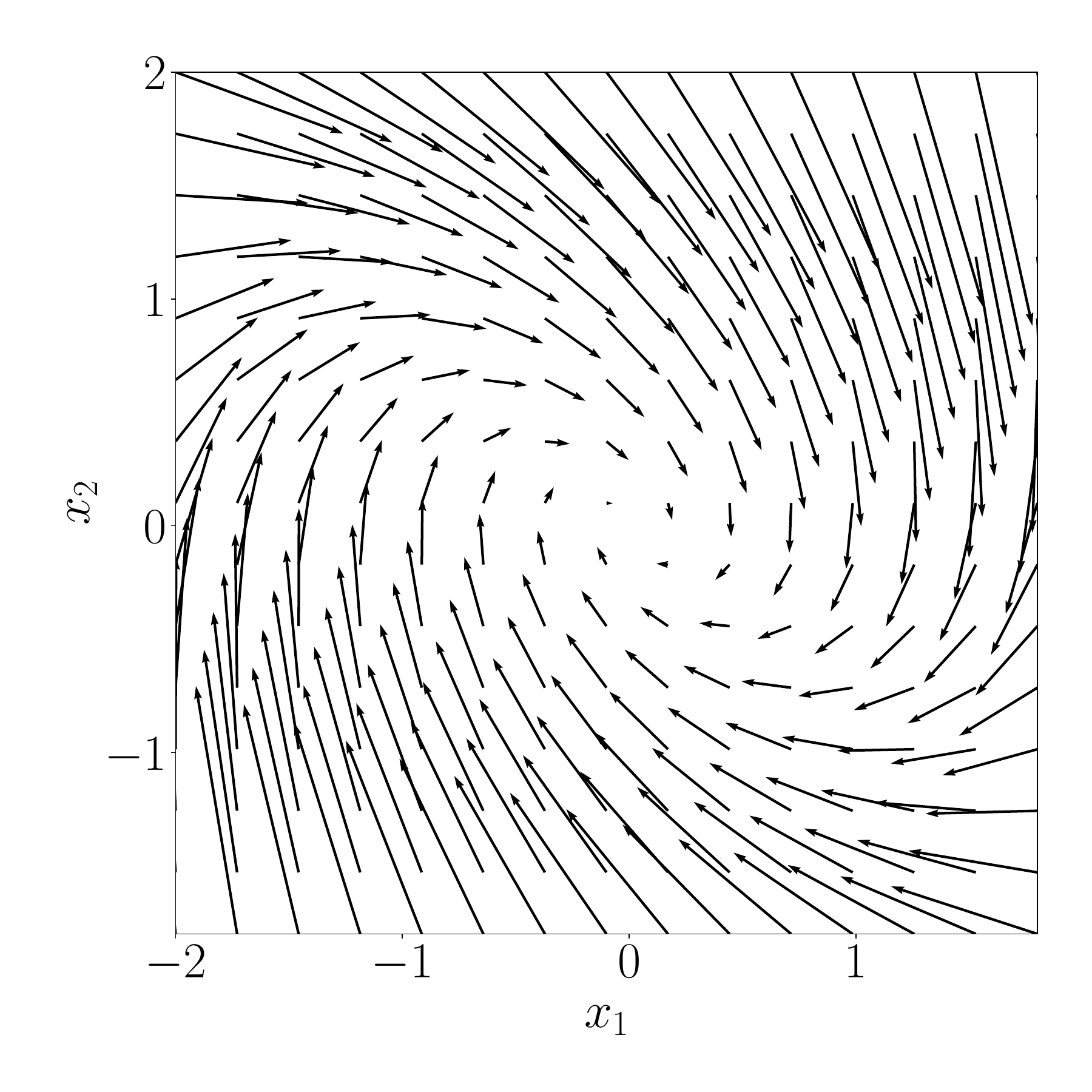}
        \caption{NN Pendulum, Vector Field}
        \label{fig:nn_dynamics}
    \end{subfigure}
    \hfill
    \begin{subfigure}[t]{0.32\textwidth}
        \includegraphics[width=\linewidth]{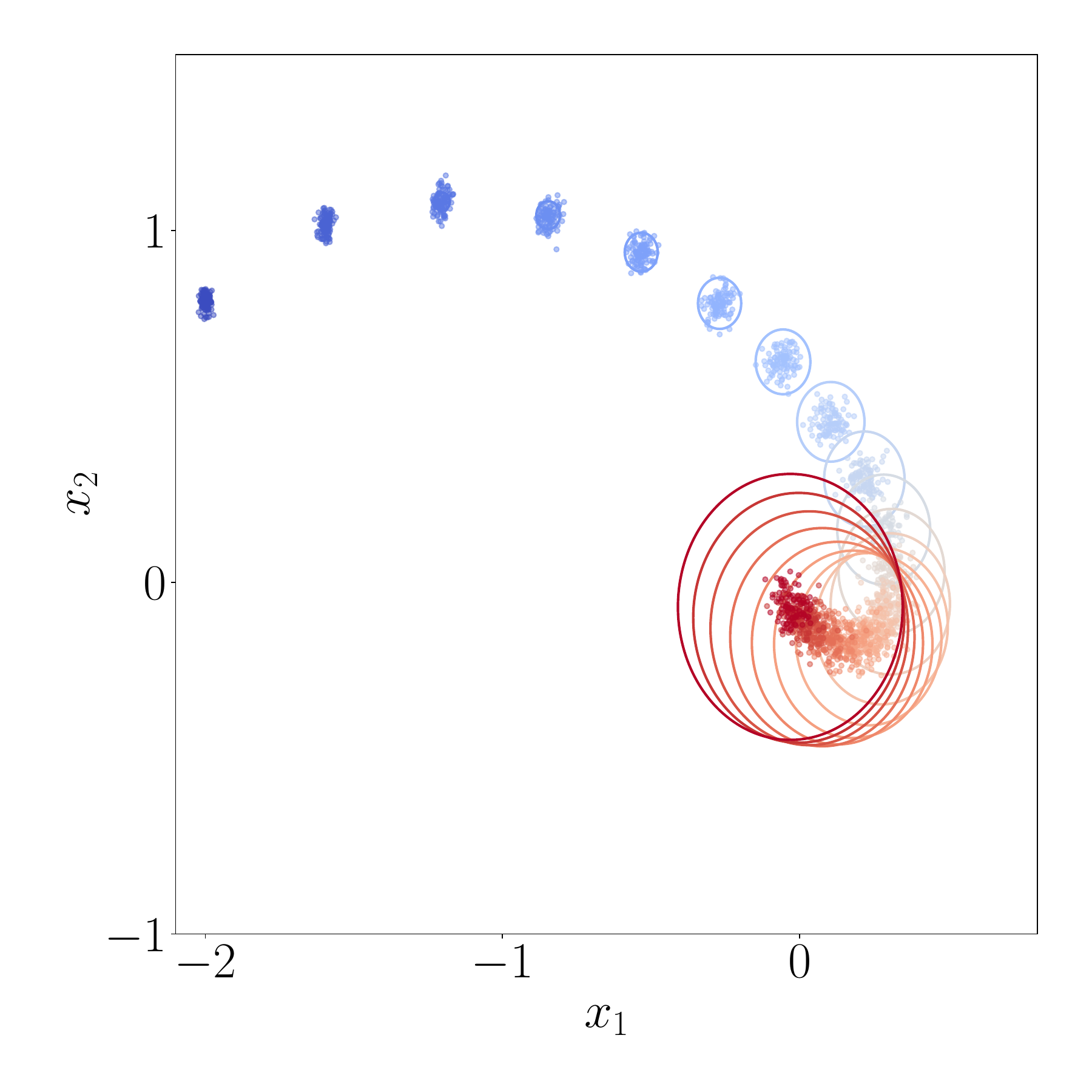}
        \caption{NN Pendulum, True Distributions}
        \label{fig:nn_true_dist}
    \end{subfigure}
    \hfill
    \begin{subfigure}[t]{0.32\textwidth}
        \includegraphics[width=\linewidth]{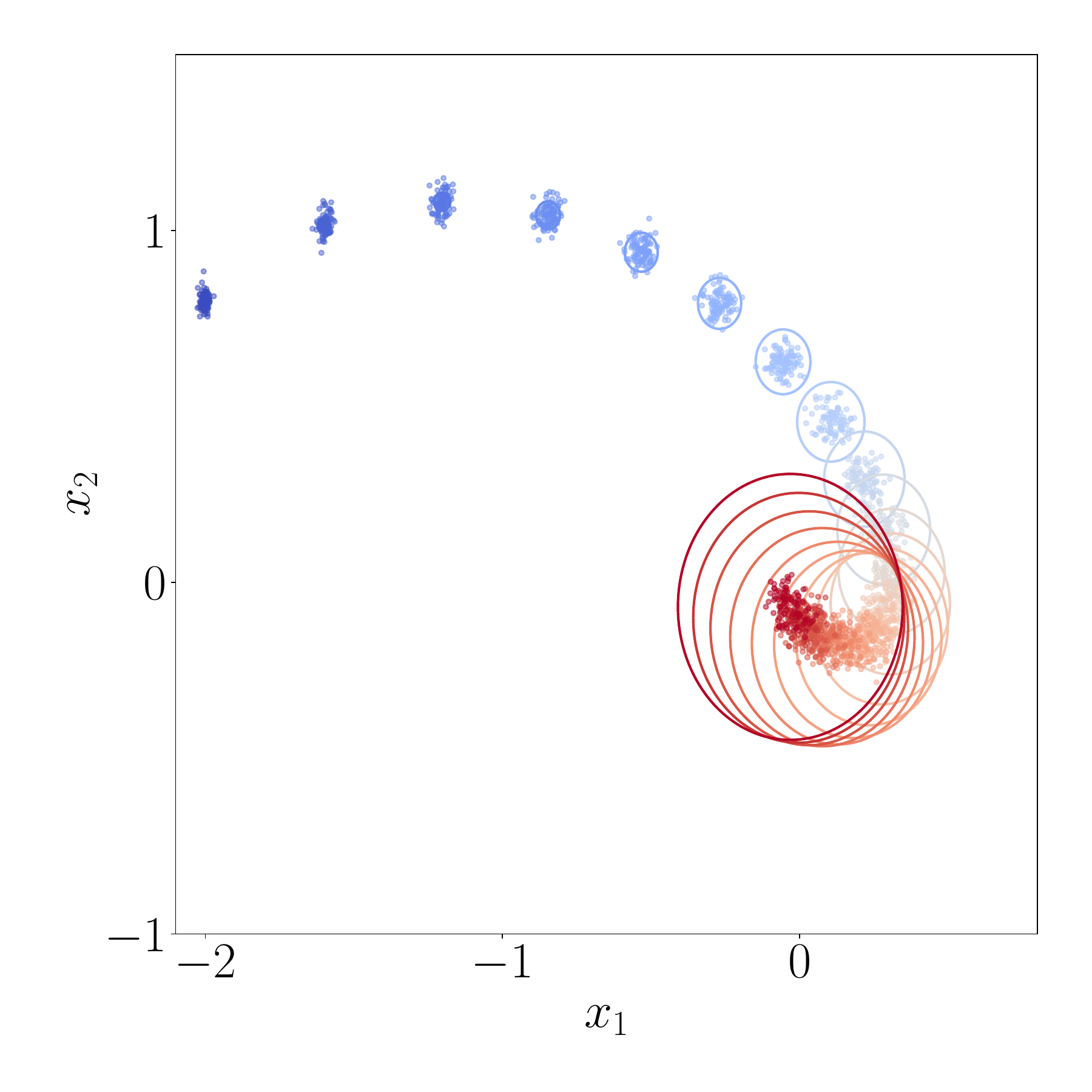}
        \caption{NN Pendulum, Approx. Distribution}
        \label{fig:nn_approx_dist}
    \end{subfigure}
  \caption{
  Vector fields (left column), samples from the true uncertainty sets \(\ambigSet_{\vx_k}\) obtained via Monte Carlo sampling (center column), and samples from the centers \(\bar\Prob_{\vx_k}\) of the ambiguity sets obtained via Algorithm~\ref{alg:propagate-ball} (right column) for different benchmarks (by row),
  evaluated over 20 time-step. 
  Sampled states are plotted as as dots, with the blue-red gradient indicating time evolution from the initial to the final time step. 
  The circles, centered at the mean of \(\bar\Prob_{\vx_k}\) with radius \(\theta_{\vx_k}\), are guaranteed to contain the mean of the true distributions. 
  The piecewise linear dynamics result from a switched system by the optimal switching strategy in \cite{gracia2025efficient}, which steers the system toward the target region while avoiding the obstacles.}
  \label{fig:exp}
\end{figure*}

In this section, we experimentally evaluate the effectiveness of Algorithm~\ref{alg:propagate-ball} in propagating uncertainty in various benchmarks.
We begin with a qualitative evaluation of the ambiguity sets computed with our framework, highlighting their structure and evolution over time and illustrating how they can be used to derive bounds in expectation.
Next, we analyze how the quantization and compression steps, as well as the uncertainty radii for the initial and noise distributions, impact the size of the resulting ambiguity sets over time\footnote{Our code is available at \url{https://github.com/sjladams/DUQviaWasserstein/tree/additive_noise}}.  
For all experiments, we fix \(\rho=2\). 
We consider the following benchmarks: 
a 2-D piecewise linear system defined by \(f(\vx)=\begin{cases}\mA_1\vx & \text{if} \ \evx{1}\leq 0 \\ \mA_2\vx & \text{else} \end{cases}\), with rotation matrices \(\mA_i =0.8\cdot\begin{bmatrix}\cos(\phi_i) & -\sin(\phi_i) \\ \sin(\phi_i) & \cos(\phi_i) \end{bmatrix}\) for \(i\in\{1,2\}\), and angles \(\phi_1=\frac{1}{8}\pi\), \(\phi_2=-\frac{1}{8}\pi\), referred to as the \textit{Double Spiral} dynamics (Figure~\ref{fig:double_spiral_dynamics});
a 2-D switched system from \cite{gracia2025efficient} with five linear modes under an optimal reach-avoid strategy, resulting in \textit{piecewise linear} dynamics (Figure~\ref{fig:switched_dynamics});
a 2-D Neural Network (NN) dynamical system where \(f\) is an MLP with two hidden layers of width \(64\) and sigmoid activations trained to approximate the discrete-time dynamics of a pendulum \cite{oriolo2025nonlinearstability}, referred to as \textit{NN Pendulum} (Figure~\ref{fig:nn_dynamics});
and an instance of the 4-D \textit{Quadruple-Tank} dynamics from \cite{johansson2000quadruple}.
For all benchmarks, the initial uncertainty sets are centered at \(\bar\Prob_{\vx_0}=\nDist(\mu_\vx, \diag(\vv_\vx))\). For the 2-D benchmarks, the noise uncertainty set is centered at \(\bar\Prob_{\vomega}=\nDist(\vzero, \diag(\vv_\vomega))\), and for the 4-D Quadruple-Tank \(\bar\Prob_{\vomega}=\frac{1}{2}\nDist(\mu_\vomega, \diag(\vv_\vomega))+\frac{1}{2}\nDist(-\mu_\vomega, \diag(\vv_\vomega))\). The corresponding parameter values are listed in Table~\ref{table:exp}. 

\begin{table}[h]
\centering
\begin{tabular}{l|c} \toprule
Dynamics & Distributions parameters \\ \midrule
Double  & \(\mu_\vx=[0.1, -0.5]\),  \(\vv_\vx= 10^{-3}\cdot\vone\)  \\
 Spiral& \(\vv_\vomega=0.0001\cdot\vone\) \\ \midrule
Piecewise  & \(\vmu_\vx=[-1.8, 1.9]\) or \([1.6, -1.9]\), \(\vv_\vx=0.0005\cdot\vone\) \\
Linear& \(\vv_\vomega=[0.0009, 0.0009]\)\\ \midrule
NN & \(\vmu_\vx=[-2, 0.8]\), \(\vv_\vx=[0.0001, 0.0005]\) \\
 Pendulum& \(\vv=[0.0001, 0.0005]\) \\ \midrule
Quadruple & \(\vmu_\vx=[1.5,2.5,0.5,1.0]\), \(\vv_\vx=[0.001, 0.02,0.4,0.01]\)  \\
-Tank& \(\vmu_\vomega=\pm0.01\cdot\vone\), \(\vv_\vomega=[0.01,0.01,0.0002,0.001]\) \\ \bottomrule
\end{tabular}
\caption{
Parameter values for the center distributions of the initial and noise uncertainty sets. 
}
\label{table:specs_uncertainty_sets}
\end{table}

\subsection{Qualitative analysis of the ambiguity set evolution}
We qualitatively validate our ambiguity sets by analyzing the time evolution of both the true distributions, approximated using Monte Carlo samples, and the 2-Wasserstein ambiguity sets obtained with our framework for the 2-D benchmarks. From Figure~\ref{fig:exp}, we first visually observe that the centers of our ambiguity sets closely resemble the true distributions. 
This is expected, as the centers of our ambiguity sets are obtained by propagating distribution \(\bar\Prob_{\vx_0}\), which is known to lie within the true set of possible initial distributions through the system dynamics. Importantly, the bimodality of the Double Spiral in its converging sequence and the transient bimodal structure in the third time step of both paths of the Piecewise Linear system, underscore the effectiveness of mixture-based approximations. Such multimodal behavior cannot be easily captured with usual propagation techniques such as moment matching \cite{deisenroth2011pilco}.

To illustrate the 2-Wasserstein ambiguity set obtained with Algorithm \ref{alg:propagate-ball}, we plot circles centered at the mean of \(\bar\Prob_{\vx_k}\) with radii \(\theta_{\vx_k}\). 
Since the \(\rho\)-th Wasserstein distance implies closeness in the \(\rho\)-th moment (see e.g., Lemma~2 in \cite{adams2024finite}), these circles are guaranteed to include the mean of the true distributions. 
In Figure~\ref{fig:double_spiral_true_dist}, we observe that for the contracting Double Spiral dynamics, the ambiguity sets converge to a fixed point, as expected from Proposition \ref{prop:convergence-contractive}. 
Interestingly, the position of the converging circles, which includes the means of the true bimodal true distributions, indicates that more mass is concentrated in the left mode than in the right. 
In contrast, for the NN Pendulum dynamics, we observe slowly expanding ambiguity sets. This can be attributed to the difficulty of computing tight, global-Lipschitz like, norm-linearizations for Neural Networks. Such computations require relaxing the network, which introduces conservatism and leads to expanding radii. 
Nevertheless, the ambiguity sets still provide informative insight on the closeness in expectation to the equilibrium of the true distribution expectation after 20 time steps.  For the Piecewise linear system, which includes non-contracting modes, the ambiguity sets gradually grow over time, as expected. In this case, the circles enable us to conclude that in expectation none of the true distributions enters the obstacle regions before reaching the target regions.

\subsection{Parametric analysis}
We start our analysis from the upper part of Table~\ref{table:exp}, which evaluates how the hyper-parameters of Algorithm~\ref{alg:propagate-ball}, that is, the quantization size \(|\bsC|\) and the number post-compression components \(N\), influence the radius \(\theta_{\vx_k}\) of the 2-Wasserstein ambiguity sets after 20 time steps. 
As expected, we observe a monotonic decrease in the ambiguity radius as \(|\bsC|\) or \(N\) increases, which confirms that finer quantization and richer mixture approximations lead to tighter ambiguity sets. 
The relative minor improvement when increasing the signature size from \(100\) to \(1000\) suggests that a relatively small number of \(100\) quantization points suffices for accurate approximations. This also holds for the higher dimensional Quadruple Tank benchmark, indicating that our approach is able to obtain tight bounds also for higher dimensional systems, by only placing locations in regions with high probability mass.  
While one might expect only marginal gains from increasing the number of components from \(5\) to \(10\), given the unimodal and bimodal behavior of the systems in Figure~\ref{fig:exp}, we observe a consistent decrease in radius. 
This effect arises because the compression step is applied to the discrete distribution \(\DiscProb_{k+1}\) in line 4 of Algorithm~\ref{alg:propagate-ball}, effectively reducing its support size and introducing approximation error.  
In the lower part of Table~\ref{table:exp}, we analyze how the ambiguity sets evolve for different values of the initial and noise uncertainty radii \(\theta_{\vx_0}\) and \(\theta_{\vomega}\). As expected, the final radius \(\theta_{\vx_k}\) after 20 time steps is largely determined by the noise uncertainty. The exponential accumulation of \(\theta_{\vomega}\) over time also explains the sharp increase in ambiguity radius when it increases from \(0.01\) to \(0.1\).

\begin{table}[h]
\centering
\begin{tabular}{ll|ccc} \toprule
\(|\bsC|\) & \(N\) & Quadruple Tank & Double Spiral & NN Pendulum \\ \midrule
10 & 1     & 0.524 & 0.115 & 0.554 \\
          & 5     & 0.437 & 0.110 & 0.492 \\
          & 10    & 0.400 & 0.105 & 0.447 \\ \midrule
100 & 1     & 0.456 & 0.103 & 0.466 \\ 
          & 5     & 0.393 & 0.099 & 0.423 \\
          & 10    & 0.354 & 0.096 & 0.391 \\ \midrule
1000 & 1     & 0.418 & 0.098 & 0.424 \\ 
          & 5     & 0.369 & 0.095 & 0.392 \\
          & 10    & 0.325 & 0.093 & 0.355 \\ \midrule
\(\theta_{\vx_0}\) & \(\theta_{\vomega}\) &&&\\ \midrule
0.001 & 0.001 & 0.358 & 0.054 & 0.423 \\
      & 0.01  & 0.393 & 0.099 & 0.610 \\
      & 0.1   & 0.748 & 0.544 & 2.488 \\ \midrule
0.01 & 0.001 & 0.358 & 0.054 & 0.433 \\
      & 0.01  & 0.393 & 0.099 & 0.620 \\
      & 0.1   & 0.748 & 0.544 & 2.497 \\ \midrule
0.1 & 0.001 & 0.358 & 0.056 & 0.531 \\
      & 0.01  & 0.394 & 0.100 & 0.718 \\
      & 0.1   & 0.748 & 0.545 & 2.596 \\ \bottomrule
\end{tabular}
\caption{Radii of the 2-Wasserstein ambiguity balls, \(\theta_{\vx_k}\), obtained via Algorithm~\ref{alg:propagate-ball} after \(20\) time steps. The table has two parts: the upper part show results for different quantization (\(|\bsC|\)) and compression (\(N\)) sizes, with fixed uncertainty radii \(\theta_{\vx_0}=\theta_{\vomega}=0.001\) for the NN Pendulum and \(\theta_{\vx_0}=\theta_{\vomega}=0.01\) else; the lower part shows results for varying \(\theta_{\vx_0}\) and \(\theta_{\omega}\), with \(|\bsC|=100\) and \(N=10\).
}
\label{table:exp}
\end{table}


\section{Conclusion}
We introduced a novel framework to formally propagate uncertainty in non-linear stochastic dynamical systems with additive noise. Our approach relies on enveloping the uncertainty in the system in tractable convex ambiguity sets of probability distributions, described as \(\rho\)-Wasserstein ambiguity sets. By allowing for uncertain initial state and noise distributions, our formulation encompasses a large class of modern control problems, such as Neural Network dynamical systems or data-driven systems. On a set of numerical experiments based on benchmarks from the control community, we demonstrated the efficacy of our framework. As future research, we see at least two directions for further improvement of the current method: i) alternative methods for compression of Gaussian mixtures, needed to guarantee scalability to the framework, and ii) less conservative ways to compute the linearization coefficients in Eqn \eqref{eq:norm-linearization}.


\section{Proofs}\label{sec:proofs}
\paragraph{Proof of Proposition \ref{prop:theor_bound_theta}}
We will prove this Proposition by induction. First, we know that $\ambigSet_{\vx_{0}} \subseteq \ball_{\theta_{x_0}}(\bar\Prob_{\vx_{0}})$ since $\ambigSet_{\vx_{0}} = \ball_{\theta_{x_0}}(\bar\Prob_{\vx_{0}})$, which is our base case. Now, for $k\geq 0$ assume $\ambigSet_{\vx_{k}} \subseteq \ball_{\theta_{x_k}}(\bar\Prob_{\vx_{k}})$. Then, for any $\ProbQ \in \ambigSet_{\vx_{k+1}}$,
\begin{align*}
    \wasserstein_\rho(\ProbQ, \bar\Prob_{\vx_{k+1}}) &\leq \sup_{\ProbQ \in \ambigSet_{x_{k+1}}}\wasserstein_\rho(\ProbQ, \bar\Prob_{\vx_{k+1}}) \\
    &= \sup_{\substack{
            \Prob\in \ambigSet_{\vx_k}, \\ \Prob_\omega \in \ball_{\theta_\omega}(\bar\Prob_\omega)}}\wasserstein_\rho(f\#\Prob\ast\Prob_\omega, \bar\Prob_{\vx_{k+1}}) \\
            &\leq \sup_{\substack{
            \Prob\in \ball_{\theta_{x_k}}(\bar\Prob_{x_k}), \\ \Prob_\omega \in \ball_{\theta_\omega}(\bar\Prob_\omega)}}\wasserstein_\rho(f\#\Prob\ast\Prob_\omega, \bar\Prob_{\vx_{k+1}})\leq \theta_{\vx_{k+1}}.
\end{align*}
Thus, $\ambigSet_{\vx_{k+1}} \subseteq \ball_{\theta_{x_{k+1}}}(\bar\Prob_{\vx_{k+1}})$, which concludes the proof.
\qed

\paragraph{Proof of Proposition \ref{prop:main-bound-result}}
First, we note that for any $\Prob, \hat\Prob, \ProbQ, \hat\ProbQ \in \sP_\rho(\sX)$:
\begin{align*}
\wasserstein_\rho(\Prob\ast\ProbQ,\hat\Prob\ast\hat\ProbQ) &\leq \wasserstein_\rho(\Prob\ast\ProbQ,\hat\Prob\ast\ProbQ) + \wasserstein_\rho(\hat\Prob\ast\ProbQ,\hat\Prob\ast\hat\ProbQ) \\
&\leq \wasserstein_\rho(\Prob,\hat\Prob) + \wasserstein_\rho(\ProbQ,\hat\ProbQ), 
\end{align*}
where the first inequality holds by the triangle inequality and the second one by Proposition 10 in \cite{aolaritei2022distributional}.
Then, for any $\Prob \in \ball_{\theta_{\vx_k}}(\bar\Prob_{\vx_k})$, $\Prob_\vomega \in \ball_{\theta_{\vomega}}(\bar\Prob_{\vomega})$, the previous result yields:
\begin{align}
&\wasserstein_\rho(f\#\Prob\ast\Prob_\vomega, \bar\Prob_{\vx_{k+1}}) \nonumber\\ 
&\qquad = \wasserstein_\rho(f\#\Prob\ast\Prob_\vomega, f\#\Delta_{\bsC_k}\#\bar\Prob_{\vx_{k}}\ast\bar\Prob_\vomega) \nonumber\\
&\qquad \leq \wasserstein_\rho(f\#\Prob, f\#\Delta_{\bsC_k}\#\bar\Prob_{\vx_{k}})+\wasserstein_\rho(\Prob_\vomega, \bar\Prob_\vomega) \nonumber\\
&\qquad \leq \wasserstein_\rho(f\#\Prob, f\#\Delta_{\bsC_k}\#\bar\Prob_{\vx_{k}})+\theta_\vomega. \label{eq:proof-sup}
\end{align}
To conclude, we apply the sup operator on both sides of Eqn \eqref{eq:proof-sup} and use Theorem 3 in \cite{figueiredo2025efficient} to the left term.
\qed

\paragraph{Proof of Proposition \ref{prop:convergence-contractive}}
First, we note that by construction, as the coefficient pair $(\alpha_{k, \ell}, \beta_{k,\ell}) = (\sL_f, 0)$ always satisfies Eqn \eqref{eq:norm-linearization} for any point $c_{k, \ell} \in \bsC_k$, it holds that:
\begin{align*}
    &\theta_\omega + \bigg(\hat{\alpha}_k(\theta_{k}+\theta_{\Delta, k})^\rho + \sum_{\ell=1}^{|\bsC_k|}\bar\Prob_{\vx_k}(\sR_{k, \ell})\beta_{k, \ell}\bigg)^\frac{1}{\rho} \\ 
    &\leq   \theta_\omega + \sL_f(\theta_{k}+\theta_{\Delta, k}) \, \leq \, \theta_\omega + \sL_f(\theta_{k}+\epsilon).
\end{align*}
where the second inequality comes from the fact that $\theta_{\Delta, k} \leq \epsilon$. Now, note that the map $T(\theta) = \theta_\omega + \sL_f(\theta+\epsilon)$ is contractive for $\sL_f<1$ (i.e. $|T(\theta)-T(\Tilde\theta)| \leq |\theta-\Tilde\theta|$). Then, by the Banach fixed point theorem \cite{goebel1990topics}, for the sequence $\theta_{k+1}=T(\theta_t)$, it holds that $\lim_{k\to\infty} \theta_k = \theta^*$, where $\theta^*$ is given by
\begin{equation*}
    \theta^* = T(\theta^*) \iff \theta^* = \frac{\theta_\omega}{1-\sL_f} + \frac{\sL_f}{1-\sL_f}\epsilon,
\end{equation*}
which concludes the proof.
\qed


\section{Acknowledgments}
L.L. and E.F. are partially supported by the NWO (grant OCENW.M.22.056).

\addtolength{\textheight}{-3cm}   

\bibliographystyle{IEEEtran}
\bibliography{bibliography}


\end{document}